\documentstyle[psfig,aps,prl,twocolumn]{revtex}
\newcommand{\be}{\begin{equation}}
\newcommand{\ee}{\end{equation}}
\newcommand{\bea}{\begin{eqnarray}}
\newcommand{\eea}{\end{eqnarray}}

\begin{document}
\twocolumn[
\begin{center}
{\LARGE\bf 
High-Power Directional Emission from Microlasers with Chaotic Resonators}\\
\vspace*{0.5cm}
{\large 
Claire Gmachl\footnotemark[2], 
Federico Capasso\footnotemark[1]\footnotemark[2], 
E.~E.~Narimanov\footnotemark[3],\\
Jens U.~N{\"o}ckel\footnotemark[4], 
A.~Douglas Stone\footnotemark[3], J{\'e}r{\^o}me 
Faist\footnotemark[2]\footnotemark[4], 
Deborah L.~Sivco\footnotemark[2], and Alfred Y.~Cho\footnotemark[2]}\\
\vspace{0.5cm}
{{\em Published in} Science {\bf 280}, 1556 (1998)}
\end{center}
\vspace{1cm}
{\bf High-power and highly directional semiconductor cylinder-lasers
based 
on an optical resonator with deformed cross section are reported. In the 
favorable directions of the far-field, a power increase of up to three 
orders of magnitude over the conventional circularly symmetric lasers was 
obtained. A ``bow-tie''-shaped resonance is responsible for the improved 
performance of the lasers in the higher range of deformations, in contrast 
to ``whispering-gallery''-type modes of circular and weakly deformed lasers. 
This resonator design, although demonstrated here in midinfrared 
quantum-cascade lasers, should be applicable to any laser based on 
semiconductors or other high-refractive index materials.}\\\vspace{1cm}
]
\flushbottom
\footnotetext[1]{To whom correspondence should be addressed;
email:\\\hspace*{0.47 cm} fc@lucent.com}
\footnotetext[2]{Bell Laboratories, Lucent Technologies, 600 Mountain
\\\hspace*{0.47 cm} Avenue, Murray Hill, NJ 07974, USA}
\footnotetext[3]{Applied Physics, Yale University, P.O. Box
208284,\\\hspace*{0.47 cm} New Haven, CT 06520, USA}
\footnotetext[4]{Max-Planck-Institut f{\"u}r Physik komplexer Systeme,\\
\hspace*{0.47 cm} N{\"o}thnitzer Str.~38, D-01187 Dresden, Germany}
\footnotetext[5]{present address: Universit{\'e} de
Neuch{\^a}tel, rue Br{\'e}guet 1,\\
\hspace*{0.47 cm} CH-2000 Neuch{\^a}tel, Switzerland}
Lasers consist of two basic components. First, the active material in 
which light of a certain wavelength range is generated from an external 
energy source, such as electric current; second, the laser resonator, 
which contains the active material, provides feedback for the stimulated 
emission of light. The resonator largely influences the special features 
of the emitted light: power, beam directionality, and spectral properties, 
as well as the laser's physical features such as size and shape. 
Semiconductor lasers are the most widely used and versatile class of 
lasers. Their most common resonators are Fabry-Perot cavities, in which 
two cleaved semiconductor crystal planes act as parallel mirrors, 
reflecting the light back and forth through the active material.

There have been many attempts to improve resonator properties. In 
particular, an increase of the reflectivity of the resonator mirrors is 
highly desirable. This allows low thresholds for the onset of laser action 
and a smaller volume of active material with concomitant moderate energy 
requirements and the ability to pack the lasers in a small space.

One excellent example is the development of micro-disk semiconductor 
lasers (1). These lasers exploit total internal reflection of light to 
achieve a mirror reflectivity near unity. Micro-disk, -cylinder or 
-droplet lasers form a class of lasers based on circularly symmetric 
resonators, which lase on ``whispering-gallery modes'' of the 
electromagnetic field (2,3,4). In such a mode light circulates around the 
curved inner boundary of the resonator, reflecting from the walls of the 
resonator with an angle of incidence always greater than the critical 
angle for total internal reflection, thus remaining trapped inside the 
resonator. There are only minute losses of light caused by evanescent 
leakage (tunneling) and scattering from surface roughness. This principle 
allowed the fabrication of the world's smallest lasers (2). Besides 
possible future applications in optical computing and networking, 
micro-lasers are of strong interest for research problems of cavity 
quantum electrodynamics, such as resonator-enhanced spontaneous emission 
and threshold-less lasers (5). Small resonators may also serve as model 
systems for the study of wave phenomena in mesoscopic systems, 
particularly in the regime where motion is fully or partially chaotic. 
Recent examples are the quantum mechanics of electrons confined in 
asymmetric ``boxes'', such as quantum-dots, stadia, and quantum corrals (6), 
and asymmetric microwave cavities with their strong connection to quantum 
chaos theory (7).

However, as a serious disadvantage, the tiny ``whispering-gallery''-type 
lasers lack high output power and directional emission because of the high 
reflectivity mirrors and the circular symmetry. Attempts to improve this 
deficiency by making gratings or small indentations on the circumference 
are so far not very promising (8,9). 
\begin{figure}[bt]
\hspace*{0.8 cm}\psfig{figure=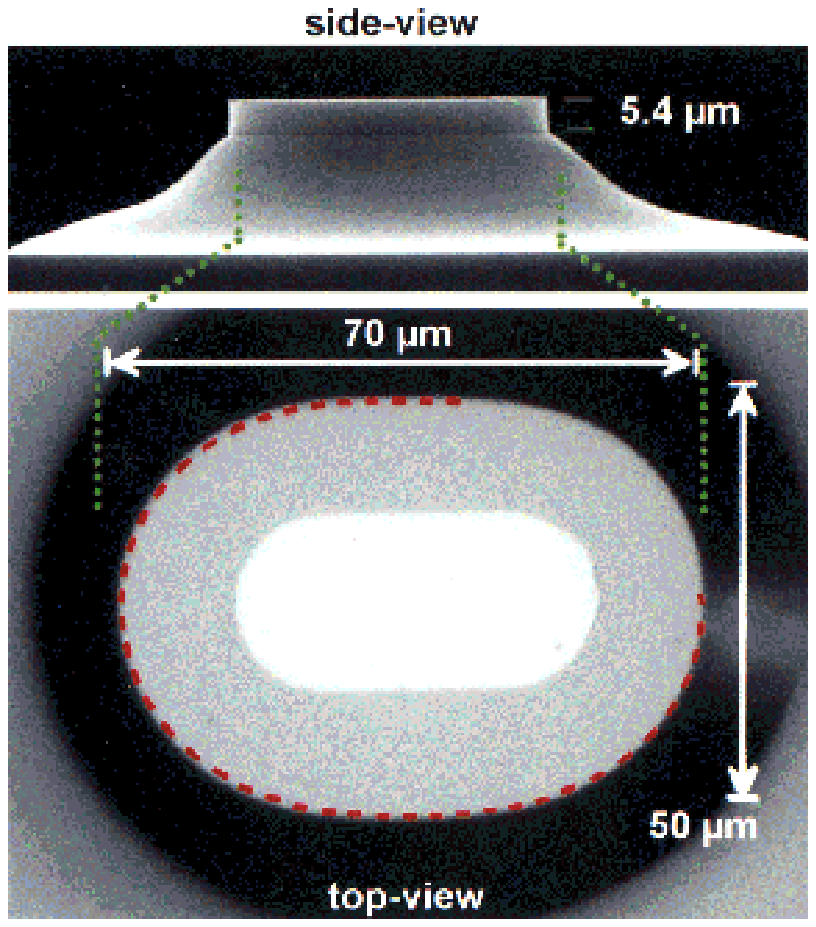,width=6 cm}

\footnotesize\noindent 
Figure 1: Scanning electron microscope image of the side and top view of a 
flattened quadrupolar shaped cylinder laser. The quadrupolar deformation 
parameter is $\epsilon\approx   0.16$. 
Side view: The laser waveguide and active material 
is entirely contained in the disk Ð with vertical side walls, total 
thickness $d = 5.39\,\mu$m Ð sitting on a sloped Indium-Phosphide pedestal. 
Light emission occurs in the plane of the disk. Top view: The top face of 
the laser is shown in medium gray, the electric contact in light gray. The 
laser boundary is very well described by an exact flattened quadrupole 
with $\epsilon = 0.16$ which is drawn as dashed red line over the circumference. 
The shape of the top electrical contact approximately parallels the laser 
shape.
\end{figure}
\noindent
We now show in
experiment and theory 
how a new resonator design which incorporates chaotic ray motion can 
dramatically increase the output power and directionality of such lasers. 
This effect is demonstrated in semiconductor quantum-cascade lasers 
emitting in the mid-infrared wavelength region (10).

Recent theoretical work has provided insight into the behavior of 
``asymmetric resonant cavities'' (ARCs), ``whispering-gallery'' resonators 
with smooth deformations from cylindrical or spherical symmetry (11-14). 
The ray dynamics in these deformed resonators is known to be either 
partially or fully chaotic in the generic case (13). The most well-studied 
example is a two-dimensional (2D) resonator with a quadrupolar deformation 
of the circular boundary, described in polar coordinates ($r,\,\phi$) by 
$r(\phi) \propto (1 +\epsilon\,\cos(2\,\phi))$, where $\epsilon$ 
is the deformation parameter. Partially 
chaotic ``whispering-gallery'' modes in these resonators have shown 
directional lasing emission in low-index materials (index of refraction 
$n < 2$, such as glass fibers or cylindrical dye jets) (12). The origin of 
the directional emission is the following (11): the deformed boundary 
causes the angle of incidence of a ray in a ``whispering-gallery'' mode to 
fluctuate in time. Eventually a ray trapped by total internal reflection 
impinges on the boundary below the critical angle and escapes by 
refraction. However, it was not appreciated in this earlier theoretical 
work that, in high-index materials, qualitatively different modes not of 
the ``whispering gallery'' type might be relevant to the lasing properties.

Here we focus on semiconductor lasers that have an effective index 
of $n \approx  3.3$ and a deformation of the boundary best described by
$r(\phi)\propto\sqrt{1+2\,\epsilon\,\cos(2\,\phi)}$, 
which we will 
refer to as a ``flattened'' quadrupole. In general, one can parameterize the 
boundary of any convex resonator by an arbitrary Fourier series for 
$r(\phi)$; 
the above parameterization is chosen for convenience, because it is simply 
analyzed and describes the actual resonator shapes quite well (see
Fig.~1). 
We show that for small deformations $\epsilon$, the basic picture of chaotic 
``whispering gallery'' orbits escaping refractively, as described above, 
still holds for the high-index semiconductor material. However, we also 
present strong experimental evidence that at larger deformations a 
different type of laser resonance emerges and is responsible for highly 
directional and high power emission. Unlike the chaotic ``whispering 
gallery'' modes of smaller deformations, these so-called ``bow-tie'' 
resonances are stable resonator modes surrounded on all sides (in phase 
space) by chaotic motion.

This new class of laser-resonators, based on the smooth deformation of a 
regularly shaped monolithic cavity, is universally applicable to 
semiconductor lasers or solid-state lasers based on high refractive index 
material. It is to some extent related to ring-lasers with resonators 
formed by assembly of several distinct flat or curved mirror surfaces, 
such as the Ti:Sapphire laser (15) or optically pumped monolithic 
solid-state lasers (16).  Nevertheless we chose the quantum cascade (QC) 
laser, as it is particularly suited for 2D `'whispering-gallery'' geometries 
as shown by a recent work on QC microdisk laser (17). It is based on a 
transition between quantized conduction band states of a cascaded 
InGaAs/InAlAs coupled quantum-well structure (intersubband transition). As 
such, the selection rule of the optical transition allows light emission 
only in the 2D plane with polarization normal to the quantum well layers; 
that is transverse magnetic (TM) polarization (18). Therefore, no light is 
lost vertical to the laser plane. Furthermore, the QC laser is a unipolar 
device based only on electron transport, unlike diode lasers. Thus, in 
contrast to most conventional semiconductor lasers, the surface can not 
cause excess unwanted non-radiative recombination of electrons and holes. 
Finally, the wavelength is comparatively large (several micrometers) and 
the material used is the well-understood InGaAs/InAlAs system. This choice 
reduces the importance of roughness (Rayleigh) scattering and makes it 
easier to fabricate complex shapes.

{\bf Device structure and experimental procedure.}
The lasers are slab-waveguide 
structures made from a ${\rm Ga}_{0.47}{\rm In}_{0.53}$As/\-Al$_{0.48}{\rm In}_{0.52}{\rm As}$ heterostructure grown 
by molecular beam epitaxy (MBE) on InP substrate. The waveguide core 
contains the QC laser active material, designed to emit light of 
$\lambda = 
5.2\,\mu$m wavelength. This active material has been used previously (19) and can 
be considered a mature and optimized design for high quality laser 
performance.

The waveguide core is sandwiched between two cladding layers (20-22). The 
entire waveguide is designed to be symmetric and such that the lasing mode 
(the lowest order TM mode) has almost no ($< 0.5 \%$) overlap with the InP 
substrate. This design prevents possible detrimental effects from light 
coupling into the substrate.

The cylinder lasers are fabricated by optical lithography and wet chemical 
etching. The quadrupolar-like shape is obtained starting from a resist 
pattern that is composed of two semi-circles connected by a rectangle 
(stadium-shape). The samples are then etched until deep mesas are obtained 
(23). Because of the smoothing action of the etchant the straight section 
of the etch mask bends toward the curved parts rendering a quadrupole-like 
shape. Figure 1 shows the top and side view of a laser with 
deformation $\epsilon\approx 0.16$.

The top view shows that the edge of the resonator follows very 
well the shape of an exact flattened quadrupole. Ohmic contacts are 
applied to the front and back surface of the lasers by using non-alloyed 
Ti/Au and Ge/Au/Ag/Au, respectively.

Several sets of samples were fabricated. The deformation parameter $\epsilon$ was 
varied in $10$ steps from $0$ to $\approx 0.2$. Two different sizes were investigated 
in order to quantify and rule out size-dependent effects, one with the 
short diameter $\approx  50\,\mu$m and the long diameter varying from $\approx 
 50\,\mu$m ($\epsilon = 0$) 
to $\approx  80\,\mu$m ($\epsilon\approx   0.2$), 
the second with the short diameter $\approx  30\,\mu$m and the 
long diameter varying from $\approx  30\,\mu$m to $\approx  50\,\mu$m (24). 
The measurements 
described below showed that effects arising from the increase in cavity 
cross-section with increasing $\epsilon$ (by less than a factor of $2$, for 
$0 \le\epsilon\le 0.2$) are negligible compared with those introduced by the 
deformation.

The measurements were mainly performed by contacting the individual 
cylinder laser with a micro-probe in a cryogenically cooled 
micro-positioner stage. To obtain the far-field pattern, the individual 
laser was mounted on a sample-holder that was rotated inside the probe 
stage. The lasers were driven with current pulses (duration $50$ ns, 
repetition rate $\approx  40$ kHz), and the light-output was measured with a cooled 
HgCdTe-detector and a lock-in technique. To improve power output and avoid 
excess current heating the data presented here were taken at $40$ to 
$100$ K 
heat sink temperature. Nevertheless the maximum pulsed operating 
temperature of the lasers is $270$ K. The spectral properties were measured 
using a Fourier transform infrared (FTIR) spectrometer.

The lasers emit light according to their symmetry into all quadrants of 
the 2D laser plane. The experimental set-up did not allow to measure the 
spatially integrated power. Therefore we collected the laser output into 
an appropriate aperture. Its center angle is varied for the acquisition of 
the far-field pattern. The light output vertical to the laser plane is 
broadened by diffraction and was measured integrated over the vertical 
extension. We introduce a polar coordinate system ($r,\,\phi$) such that 
$\phi = 0^{\circ}$ 
indicates the direction along the elongated (major) axis. 
Accordingly, $\phi = 
90^{\circ}$ denotes the direction of the compressed (minor) axis. Hence a 
measurement taken at $\phi = 0^{\circ}$ has the detector facing one point of highest 
curvature of the deformed laser.

{\bf Power output and beam directionality. }
The deformed cylinder lasers 
provided both a dramatic increase of the emitted power and directionality 
(Figs. 2 and 3).

Light output measurements for various lasers as a function of their 
deformation $\epsilon$ are shown in Fig.~2A. To generate this plot, 
the maximum obtainable peak power was recorded for each laser by 
optimizing the pulsed drive current (25). The collecting aperture 
(slit width corresponding to 
$15^{\circ}$) was oriented around $\phi = 0^{\circ}$. Note that this set-up precludes the 
observation of any changes in the far-field directionality with 
deformation. Similar measurements were performed with the aperture 
oriented around $\phi = 45^{\circ}$ and $\phi = 90^{\circ}$. The striking result is the strong 
(quasi-exponential) increase of the collected optical power with 
deformation. For the largest deformation under consideration 
($\epsilon \approx  0.2$), a 
power increase of a factor of $\approx 50$ with respect to the circular case is 
observed. Figure 2A shows a representative measurement taken at $\phi = 
0^{\circ}$.

The absolute output power was measured in some highly deformed laser 
devices by bonding and mounting them in a calibrated set-up usually used 
with Fabry-Perot type lasers. One example, obtained from a laser with 
$\epsilon\approx   
0.2$, is shown in Fig.~2B. The light-collecting aperture was increased to 
its maximum size, and the sample was tilted to detect roughly the optical 
power in an angle from $+ 40^{\circ}$ to $+ 100^{\circ}$. The choice of this aperture, which 
exploits far-field anisotropy, will become clear below. A peak output 
power of $\approx  10$ mW at $100$ K was obtained. This value is approximately three 
orders of magnitude greater than that obtained from the non-deformed 
(circular cylindrical) laser or previous conventional circular QC-disk 
lasers (17). For a weakly deformed laser ($\epsilon = 0.06$) we estimate a peak 
power output of $\approx 50\,\mu$W (when measured with comparable collection 
efficiency as the laser of Fig.~2B). 

A quasi-exponential increase of the collected power with increasing 
deformation (similar to the one shown in Fig.~2A) has been measured in 
numerous sets of lasers of various - flattened and less flattened - 
quadrupolar shapes and sizes, and with various orientations of the 
aperture. Thus it appears that the power increase is a reliable, universal 
effect. However, the increase in output power per unit angle is rather 
closely entangled with the actual variation of the far-field pattern with 
deformation. In fact, in our lasers the power increase with deformation 
results from the lasing of different types of modes in different ranges of 
$\epsilon$. There is a cross-over at intermediate deformations ($\epsilon \approx 
 0.12$) from 
emission via ``whispering-gallery'' modes, which dominates at smaller
\begin{figure}[bt]
\psfig{figure=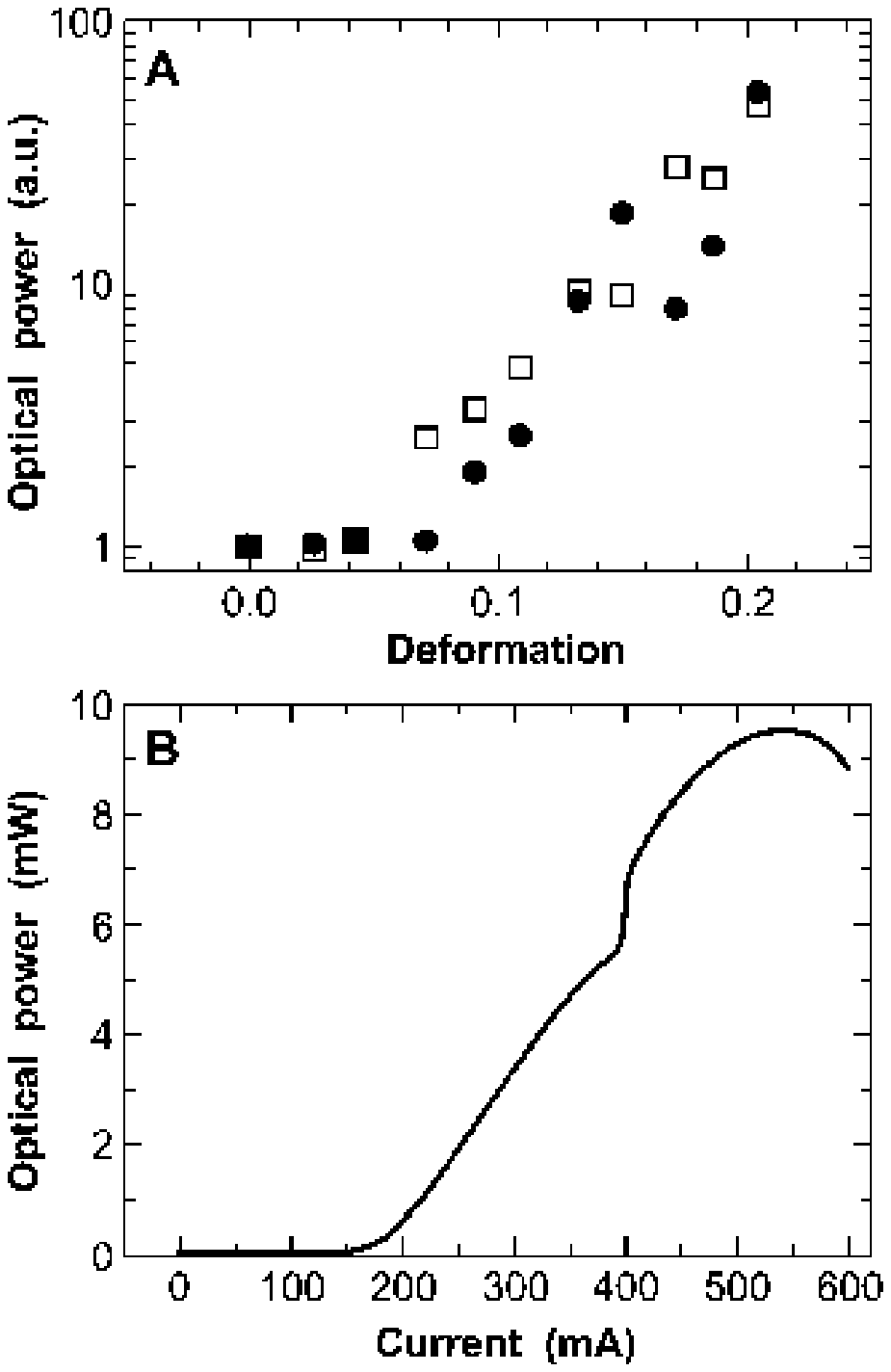,width=8.5 cm}

\footnotesize\noindent 
Figure 2: ({\bf A}) Maximum peak optical power from various lasers as function of 
their quadrupolar deformation parameter $\epsilon$. The aperture with width 
$15^{\circ}$ is 
centered around zero degree. Data from two independent sets of lasers are 
given. The power output is normalized to the power of the respective 
circular cylinder laser. The scatter of the data is due to the coarseness 
of the mode spectra (single or few versus multiple modes lasing). 
({\bf B}) Light output-to-current characteristics of a quadrupolar cylinder 
laser with deformation $\epsilon\approx   0.2$. The collecting aperture ranges 
from $+ 40^{\circ}$
to $+ 100^{\circ}$ (the polar coordinate system is described in the main text and 
in Fig.~3C). The kink around $400$ mA indicates the onset of a second lasing 
mode. The measurement has been performed at $100$ K heat sink temperature. 
The lasers have been tested up to $270$ K.
\end{figure}
\noindent
deformations, to laser emission from ``bow-tie'' modes which do not exist 
below $\epsilon \approx  0.10$ but dominate the high deformation regime.

In addition to the strong increase in power output, the deformed lasers 
can also provide strong directionality. The results of the far-field 
measurements are summarized in Fig.~3A and 3C. As expected, the circular 
cylinder laser displays no directionality of the emission. At small 
deformations ($\epsilon \le 0.10$), the far-field is only weakly structured with an 
increased emission in direction of the minor axis compared to the major 
axis. Figure 3A shows the increase of the output power with $\epsilon$, collected 
around $0^{\circ}$ and $90^{\circ}$. Both curves rise approximately exponentially, as 
discussed in the previous section, but ``faster'' for $\phi = 90^{\circ}$; in this case 
the exponent is increased by a factor of $\approx 2$ with respect to the $\phi = 
0^{\circ}$ case (26).

This observation is consistent with the expected behavior of deformed 
``whispering gallery'' modes with an average angle of incidence near the 
critical angle defined by $\sin(\chi_c) = 1/n$, where $n = 3.3$ is the effective 
refractive index of the laser waveguide.  At zero deformation such a mode 
has a conserved angle of incidence and emits isotropically and uniformly 
via evanescent leakage from all points at the boundary (neglecting 
disorder effects, such as surface roughness scattering).  However when the 
boundary is deformed (11) the angle of incidence of a ray associated with 
a lasing mode fluctuates and (at these deformations) is most likely to 
collide with the boundary below the critical angle of incidence at or near 
a location of high curvature ($\phi = 0^{\circ}, \,180^{\circ}$). Fig.~3B shows the calculated 
intensity pattern (the modulus squared of the electric field) for a 
typical ``whispering-gallery'' mode in a deformed cylinder laser with 
$\epsilon = 0.06$ (the calculational technique will be discussed below). The 
pattern shows clearly the enhanced emission intensity in the near-field in 
the vicinity of ($\phi = 0^{\circ},\, 180^{\circ}$).  
The experiments are sensitive to the 
far-field intensity distribution, which depends also on the angle of 
refraction at the points of high curvature.  Both the ray and wave 
calculations discussed below indicate that at this deformation all 
``whispering gallery'' modes with high output coupling have a minimum in 
emission intensity in the far-field around $\phi = 0^{\circ}$ and enhanced emission 
between $45^{\circ}$ and $90^{\circ}$.  
The observed experimental intensity pattern has this 
general trend (see Fig.~3A), but a fully angle- and mode-resolved 
measurement of the far-field pattern and a detailed comparison with theory 
is difficult due to the generally low optical power and the many modes 
which contribute to the laser signal in this regime of deformations. (A 
detailed discussion of the spectral properties is given in the next 
section.)

At higher deformations ($\epsilon\ge 0.14$) we detect a much stronger and 
qualitatively different directionality. Figure 3C shows the actual 
angle-resolved far-field pattern (one quadrant) of one circular and two 
deformed lasers. For the laser displayed in Fig.~1, we obtain a power 
increase by a factor of $30$ into an emission angle of $\phi = 
42^{\circ}$ compared to $\phi
= 0^{\circ}$. The angular width of this directional emission is $\approx  
23^{\circ}$. Around $0^{\circ}$ 
we observe a clear minimum of the emission, and a smooth sloping plateau 
towards $90^{\circ}$.

At these large deformations, a typical ray characterizing a 
``whispering-gallery'' mode escapes in less than $10$ collisions with the 
boundary as discussed in the theory section. This ray escape is now 
approximately isotropic and would seem unlikely to lead to the increased 
emission anisotropy observed experimentally. Because the general ray 
motion is furthermore highly chaotic in most of the phase-space, the only 
plausible scenario for generating 
\onecolumn
\begin{figure}[bt]
\hbox{\hspace{0.2cm}\psfig{figure=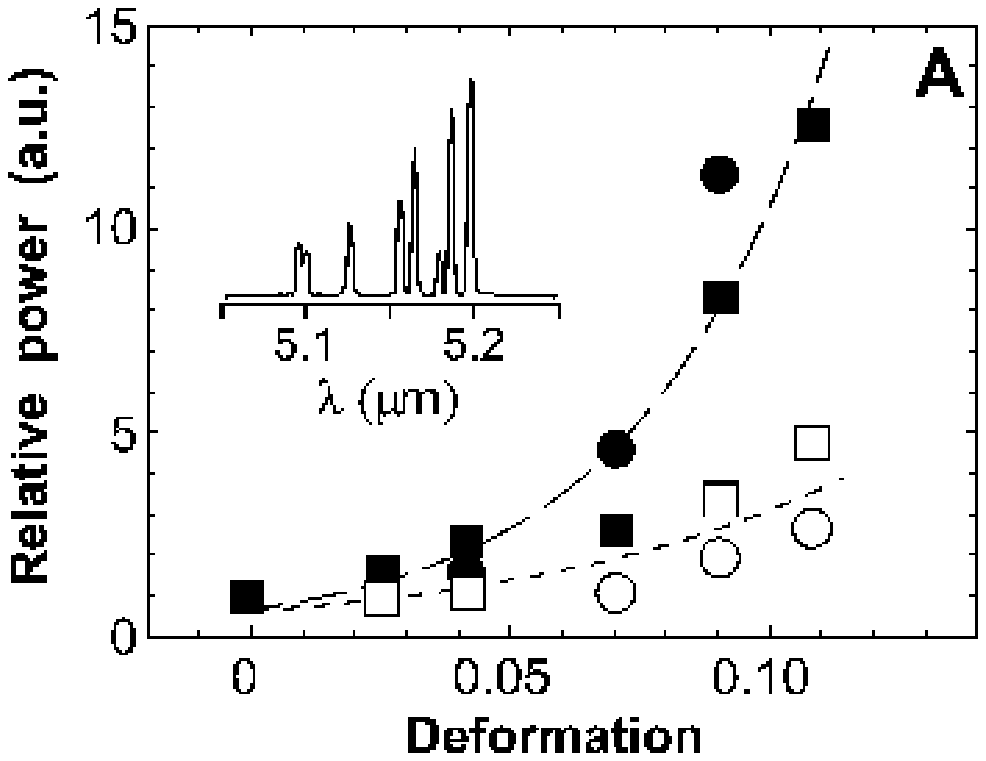,width=9 cm}\hspace*{0.4 cm}
\psfig{figure=figures/wgwave.EPSF,width=7 cm}}\vspace*{0.1 cm}
\hbox{\hspace{0.2cm}\psfig{figure=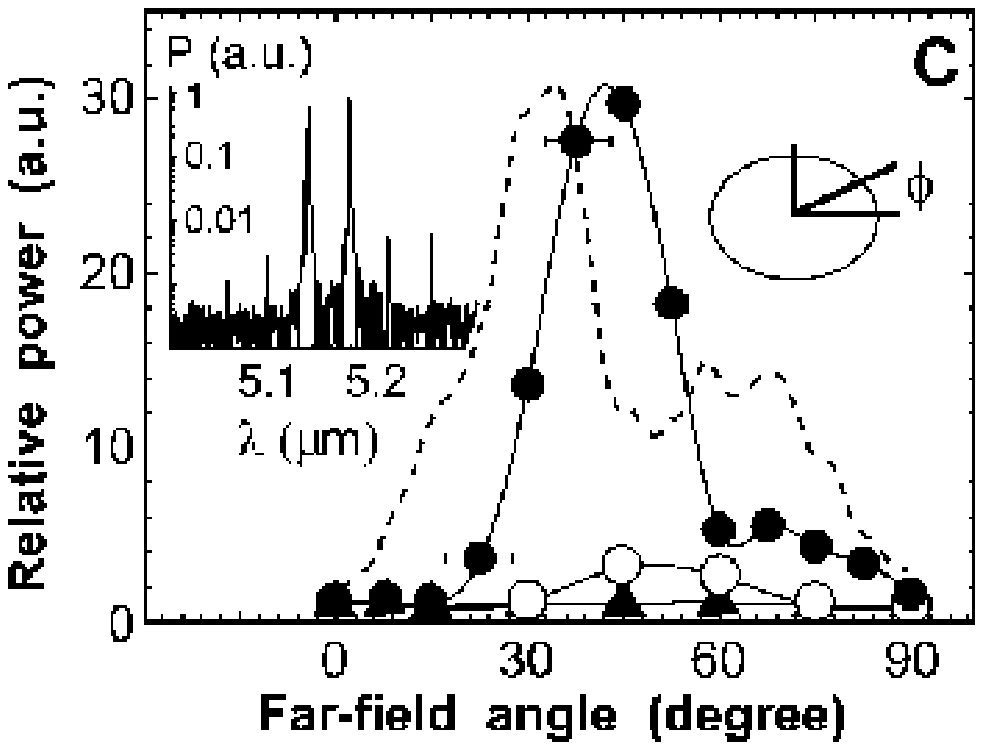,width=9 cm}\hspace*{0.4 cm}
\psfig{figure=figures/bowwave.EPSF,width=7 cm}}\vspace*{0.2cm}

\footnotesize\noindent 
Figure 3: ({\bf A}) Peak output power of different lasers as a function of 
deformation. The power is collected around $0^{\circ}$ (open symbols) and 
$90^{\circ}$ 
(filled symbols) with a width of the fixed aperture of $15^{\circ}$. Two 
independent sets of lasers are presented for each orientation of the 
aperture. Both curves rise approximately exponentially, as indicated by 
the dashed line-fit (26).
Inset: Spectrum in linear scale obtained near peak optical power from a 
cylinder laser with low deformation ($\epsilon\approx 0.04$). The close mode spacing 
observed in the spectrum is a result of several lasing 
whispering-gallery-type modes. The displayed linewidth is limited by the 
experimental set-up and data acquisition system.
({\bf B}) False color representation of the radiation intensity pattern of a 
chaotic ``whispering-gallery'' mode for a deformed cylinder with 
$\epsilon = 0.06$ 
and length of the minor axis of $50\,\mu$m. Red indicates high intensity, dark 
blue minimum intensity on a logarithmic intensity scale. The computational 
technique is explained in the theory section of the main text.
({\bf C}) Symbols indicate the measured angle-resolved far-field pattern (one 
quadrant) of a circular 
(\psfig{figure=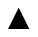}) 
and two deformed lasers with $\epsilon = 
0.14$ (\Large$\circ$\footnotesize) and $\epsilon = 0.16$ (\Large
$\bullet$\footnotesize). 
The measurements presented 
here have been taken at a constant current level, at which the (deformed) 
lasers displayed pure single mode emission. However, the far-field shows 
qualitatively the same characteristic directionality at a current level 
corresponding to peak optical power. The data-sets are normalized to the 
value measured at zero degree. The data points are connected by splines 
(solid lines) for clarity. 
The dashed line is the far-field intensity pattern associated with the 
bow-tie mode shown in Fig.~3D, averaged over the experimental aperture. 
The calculation has been scaled to match the peak emission at $\approx  
45^{\circ}$. The 
exact angular position of this maximum should be quite sensitive to the 
precise shape of the boundary near the ``bow-tie'' impact points, and we 
attribute the off-set between the measured and calculated peak positions 
primarily to the small deviation between our model and the actual shape 
and some uncertainty in the precise measurement of the angle. Furthermore, 
at present, we do not have a full understanding of the discrepancy in the 
intensities of the secondary peak between calculation and experiment.
({\bf Left inset}) Logarithmic plot of the measured spectrum at maximum power 
(power $P$ versus wavelength $\lambda$) of a laser with 
$\epsilon \approx  0.16$. Six equally spaced 
modes, with mode spacing $\Delta\lambda = 40.4$ nm, are observed. This
mode separation 
is in good agreement with the value of $39.5$ nm calculated for a ``bow-tie'' 
orbit corresponding to the calculated intensity pattern of Fig.~3D.
({\bf Right inset}) The polar coordinate system is oriented such that $\phi = 
0^{\circ}$ 
indicates the direction along the elongated (major) axis, and $\phi= 
90^{\circ} $
denotes the direction of the compressed (minor) axis.
({\bf D}) False color representation of the intensity pattern of a 
``bow-tie''-mode for $\epsilon = 0.15$ and length of the minor axis of 
$50\,\mu$m. Note 
that the cross-over to the asymptotic far-field pattern of Fig.~3C (dashed 
line) is rather slow, and that certain features such as the modulated 
intensity at $\phi = 90^{\circ}$ vanish in the far-field. The 
color scale (red - high intensity, blue - low intensity) is 
unrelated to the color scale of Fig.~3B.
\end{figure}
\twocolumn
\noindent 
directional emission is for the lasing 
modes to be associated with the small regions of stable, regular motion 
that still remain.

For the range of deformations $\epsilon\approx   0.12$ to $0.23$, 
there exist only two such 
regions. The first is the vicinity of the basic diametral orbit running 
along the minor axis of the resonator. The associated modes are the 
transverse modes of the stable, curved mirror Fabry-Perot resonator. 
However, these modes correspond to normal incidence at the boundary and, 
as such, would result in a peak emission at $90^{\circ}$ in the far-field, much in 
contrast to the observation. Furthermore, the low reflectivity of the 
boundary at normal incidence combined with the short length of the minor 
axis result in a too high threshold for laser action.

The second region is in the vicinity of the stable four-bounce periodic 
orbit with the shape of a ``bow-tie'' in real space. The intensity pattern 
of a representative ``bow-tie'' mode is shown in Fig.~3D. This orbit comes 
into existence by bifurcation from the diametral orbit at $\epsilon\approx   0.10$ 
and has 
four equal (in absolute value) angles of incidence on the boundary. At $\epsilon\approx   
0.12$ this angle $\approx  12.5^{\circ}$ and it is well below the critical angle, but as 
the deformation increases to $\epsilon\approx   0.15$, this angle increases to 
approximately the critical angle, $\chi_c \approx  17.5^{\circ}$. This 
change results in a 
sufficiently high reflectivity of the boundary to allow for laser action. 
For $\epsilon$ of $0.125,\, 0.14$, and $0.15$, values of the reflectivity 
of $0.45,\, 0.59$, 
and $0.72$ are calculated, respectively. In fact, this increase in 
reflectivity with deformation should lead to a reduction of the laser 
threshold.

When the radiation intensity pattern of a ``bow-tie'' mode is averaged 
according to the experimental conditions, we find reasonable agreement 
between the experimental and theoretical far-field directionality 
(Fig.~3C). We conclude that the laser emission at high deformations originates 
from the newly observed ``bow-tie'' modes. The spectral properties of the 
emission provide further confirmation of this fact, as discussed in the 
following section.

The ``bow-tie'' orbit is just one of several orbits that move around the 
minor axis in a librational motion (that is without a fixed sense of 
rotation) as opposed to the rotational motion of conventional ``whispering 
gallery'' orbits. With higher index of refraction or different shape 
deformations, modes associated with other librational orbits may be 
relevant to lasing; hence we will refer to the ``bow-tie'' as one of a class 
of ``librational'' modes. 

In general one would expect that the threshold current density $J_{th}$ should 
have a minimum for $\epsilon = 0$ (circular case), increase with deformation until 
$\epsilon \approx 0.1$, due to the increase in outcoupling loss, and then decrease due to 
the gradually increasing reflectivity of the ``bow-tie'' modes. In the range 
of $\epsilon \approx 0.12$ to $0.2$, the observed decrease in $J_{th}$ from $\approx 
 5$ kAcm$^{-2}$ to 
$4$ kAcm$^{-2}$ is consistent with this expectation. However, in the ``whispering 
gallery'' range of deformations $\epsilon = 0$ to 
$\epsilon\approx 0.08$ the measured decrease in 
$J_{th}$ from $\approx  7$ kAcm$^{-2}$ to $6$ kAcm$^{-2}$ is in contrast with 
the expectations. 
There are several issues that complicate the interpretation of the 
threshold data. First, there is a finite lateral current spreading 
resistance, effectively reducing the current density towards the edge of 
the disk outside the contact region. Second, the mode confinement factor 
within the active region is expected to be reduced in the outermost parts 
of the waveguide due to the true three-dimensional nature of the 
waveguide, increasing locally the threshold current density. As a 
consequence, the actual threshold current density for a given mode depends 
on its spatial distribution within the resonator. Our experimental
$J_{th}$, 
on 
the contrary, is always calculated by dividing the value of the threshold 
current by the geometrical area of the actual device. Finally lowering of 
the threshold current density will in general also lead to an increase in 
the maximum output power due to the higher available range of drive 
currents. 

The threshold current density of the QC-laser is given by $J_{th} = 
(\alpha_{wav} + 
\alpha_{out})/g_{\Gamma}$, with $\alpha_{wav}$ 
being the waveguide loss, $g_{\Gamma}$ the average modal gain 
coefficient, and $\alpha_{out}$, the outcoupling loss, which strongly depends on the 
distributed reflectivity of the boundary, which in turn depends strongly 
on the deformation $\epsilon$, and effective length of the resonator. From the 
laser threshold and the computed value of $g_{\Gamma}$ ($6.72 
\quad{\rm at}\quad
10^{-3} \,{\rm cmA}^{-1}$), the 
quality-factor (``$Q$-value'') can be calculated as $Q = (2\,\pi\,n)/
[\lambda\,(\alpha_{wav} + \alpha_{out})]$, 
where $n$ is the effective refractive index and $\lambda$ the wavelength. 
The above cited threshold current densities then result in $Q$-values 
ranging from $\approx 850$ to $1500$. It should be noted that waveguide losses are 
usually dominant in QC lasers due to the high doping levels which increase 
free carrier absorption.

{\bf Spectral properties.} 
Besides the increase in output power and 
directionality, the increasing deformation also influences the spectral 
properties of the lasers. These further confirm the existence of two 
different regimes, manifested so clearly in the different types of 
far-field patterns.

At low deformations we obtain a complex, dense modal spectrum. The lasers 
are multiple-mode starting from threshold, with close mode spacings, and 
show up to $10$ almost equally strong modes at the maximum optical power 
(see inset to Fig.~3A). This close mode spacing cannot be understood from 
one fundamental set of longitudinal (azimuthal) whispering gallery modes 
only since fitting an integer number of wavelengths along a single closed 
ray path would result in a regular comb of modes with significantly larger 
spacings. We therefore attribute the spectrum to the lasing of several 
different longitudinal (``azimuthal'') and transverse (``radial'') modes.

At large deformations the lasers are single-mode until approximately twice 
the threshold current, and show at most two to three strong modes at 
maximum power. The onset of additional modes is accompanied by a kink in 
the light output-current characteristic; one can be seen in Fig.~2B. The 
cross-over between the two spectrally characteristic regimes again occurs 
around $\epsilon \approx  0.12$.

The multiple mode behavior of the highly deformed lasers is consistent 
with the emission from ``bow-tie'' modes. The logarithmic plot of a spectrum 
in this regime (inset Fig.~3C) reveals six equally spaced modes, with mode 
spacing $\Delta\lambda = 40.4$ nm. The expected theoretical value is calculated 
assuming that adjacent modes differ by one wavelength along the 
path-length of the ``bow-tie''. This analysis yields a spacing of $39.5$ nm, 
in excellent agreement with the experiment, considering the uncertainty in 
the effective refractive index.

The ``bow-tie'' modes can easily be distinguished from transverse modes of 
the diametral curved mirror Fabry-Perot resonator along the minor laser 
axis (length $L$). As noted above, they originate from a period-doubling 
bifurcation of the latter, as will be discussed below in more detail, 
leading to approximately twice the optical path length. As such, the 
``bow-tie'' mode spectrum displays approximately half the mode spacing one 
would expect of the standard Fabry-Perot modes, $\Delta\lambda = 
\lambda^2/(2\,n\,L) \approx  
82$ nm.

To summarize, the experimental data show that imposing a flattened 
quadrupolar deformation onto semiconductor micro-lasers substantially 
improves their power output and directionality. In the favorable 
directions of the far-field a power increase of up to three orders of 
magnitude was obtained. This dramatic result could be achieved by 
exploiting the complex ray dynamics - first for chaotic 
``whispering-gallery'' modes, then for ``bow-tie'' modes - of the deformed 
resonators. An in-depth theoretical discussion of the subject is given in 
the following section.

{\bf Theory.} 
The intensity patterns shown in Fig.~3, B and D, were obtained by 
numerical solution of the Helmholtz equation for the TM polarization 
resonances at $\lambda\approx 5.2\,\mu$m of a deformed dielectric cylinder with the 
dimensions and index of refraction ($n = 3.3$)  corresponding to the 
experimental structures. These solutions are obtained by matching the 
internal and external electric fields and their derivatives at the surface 
of the semiconductor, along with the additional constraint that there is 
no incoming wave from infinity. The latter constraint implies that the 
wavevector must be complex, with the imaginary part giving the decay rate 
or $Q$-value of the resonance (28). 

To obtain a full theoretical understanding of these resonances it is 
helpful to divide the problem in two parts. First, we consider the 
properties of the ``bound-states'' of the system, corresponding to the 
discrete solutions which would exist if the cavity were completely closed 
and the electric field were zero outside the cavity. Then we must 
understand how these states are altered by the possibility of escape to 
infinity by refraction.

The first point is precisely the issue of understanding the solutions of 
the wave equation within a ``billiard''. This problem corresponds to a 
resonator with a mirror reflectivity exactly equal to unity. With these 
``hard-wall'' boundary conditions, the Helmholtz wave equation is identical 
to the Schr{\"o}dinger equation of quantum mechanics.

When the cross-section of the cylinder is deformed from circularity, the 
wave equation is no longer separable into three 1D differential equations 
and the solutions in the plane transverse to the cylinder axis are no 
longer specified by pairs of quantum numbers (or mode indices).

One can still obtain a numerical solution, that is by representing the 
solution in a large basis set of states and diagonalizing the resulting 
matrix equations. However, if this approach is used alone it is difficult 
to extract any physical understanding of the bound states or ``true'' 
resonances, now taking into account the electric field outside the 
resonator. In fact, the solutions shown in Figs.~3, B and D were predicted 
first by a completely different theoretical approach, before they were 
found by numerical search. This different approach, which has been 
pioneered in physics (29-31) and physical chemistry (32) during the past 
two decades, is to study the short wavelength limit of the problem (ray 
optics for the Helmholtz equation, Newtonian mechanics for Schr{\"o}dinger's 
equation) and try to develop a systematic understanding with semiclassical 
methods. The use of semiclassical methods is justified in our system 
because the wavelength of light in the material ($\approx 1.6\,\mu$m) 
is much smaller 
than any of the geometric features of the resonators. Moreover, standard 
perturbation techniques are not applicable because the deformation causes 
a shift in the resonance frequencies that is large compared to the 
resonance spacing.

When the optical wave equation is non-separable, the corresponding ray 
motion typically exhibits fully or partially chaotic dynamics, just as the 
classical limit of a non-separable Schr{\"o}dinger equation typically gives a 
chaotic classical mechanics; this sub-field has become known as ``quantum 
or wave chaos theory.'' The stationary states of these so-called ``quantum 
billiards'' have been studied extensively in this context. Here we will 
discuss the ray-optics properties of the billiards corresponding to the 
laser resonators studied above, with the goal of understanding the 
cross-over between emission from ``whispering-gallery'' to ``bow-tie'' modes 
which occurs in this system.

The relevant billiards are smooth deformations of the circular billiard. 
Initially we neglect the possibility of escape. Rays will simply propagate 
indefinitely within the billiard, satisfying the law of specular 
reflection at collisions with the boundary. When the circle is undeformed, 
angular momentum is conserved in this motion. The angle of 
incidence, $\chi$, 
is the same at each collision, and the orbit traces out an annulus bounded 
by a circular caustic of radius $R\,\sin(\chi)$, where $R$ is the radius of the 
circle. The corresponding wave solutions are the ordinary Bessel-functions 
indexed by the angular momentum quantum number.

When the boundary conditions are changed to include refraction, then rays 
incident with $\chi$ greater than the critical value $\chi_c$ given by 
$\sin(\chi_c) = 1/n$ 
will remain trapped by total internal reflection, whereas rays with 
$\sin(\chi)\le 1/n$ will rapidly escape by refraction according to Snell's law.

To illustrate the circular and deformed case in a unified manner we 
represent the ray motion in phase space using the surface of section (SOS) 
method (31,33), in which every time a ray collides with the boundary both 
the azimuthal angle ($\phi$), at which it hits, and its angle of incidence 
($\chi$) with respect to the boundary are recorded. Following an ensemble of 
hundred trajectories for $200$ bounces then gives a good picture of the 
global dynamics in phase space.

The generic behavior of smoothly deformed circular billiards in this 
representation is shown in Fig.~4, where again we neglect the possibility 
of escape in calculating the SOS. For the circle (Fig.~4A) the SOS is 
trivial, each trajectory gives a straight line corresponding to the 
conserved value of $\sin(\chi)$, except for trajectories with a chord angle 
($2\,\chi$) equal to a rational fraction, $p/q$, of $2\,\pi$. Such trajectories will 
close after $q$ bounces and are referred to as ``period-$q$'' orbits. All such 
orbits in the circle are marginally stable and exist in infinite families 
corresponding to arbitrary rotations of any one orbit in the family. 
Several period-2, period-3, and period-4 orbits are indicated in the SOS 
of Fig.~4A; the period-2 orbits, which are very important in the 
discussion below, just traverse the diameter of the circle.

In all of the SOSs in Fig.~4 we have indicated in red the horizontal line 
corresponding to the critical angle, $\sin(\chi_c) = 1/n = 0.30$. Trajectories 
which fall below that line in the closed billiard will escape from the 
semiconductor. Trajectories above stay ``forever'' trapped within the 
resonator [in this approximation, which neglects weak evanescent leakage 
or tunneling of photons (34)]. When the circle is deformed the ray 
dynamics in the billiard undergoes a transition to partially chaotic 
motion. If the deformation is smooth and the curvature of the boundary is 
always convex, it can be shown rigorously that the phase-space still has 
non-chaotic ``whispering-gallery'' modes for values of $\sin(\chi)$ sufficiently 
close to one (35).

The specific form of the deformation is unimportant for the qualitative 
physics; we use the flattened quadrupolar deformation, which describes 
well the experiment. One sees the effect of a deformation of $\epsilon = 0.06$ in 
Fig.~4B. For $\sin(\chi) > 0.7$, there remain many unbroken (continuous) curves 
traversing the full surface of section which correspond to 
``whispering-gallery'' modes that survive only slightly deformed from the 
circle. These are ``whispering gallery'' orbits of the familiar type, which 
are confined near the rim of the resonator, have a true caustic and will 
circulate in one sense indefinitely. However one also now sees the 
signature of isolated stable and unstable periodic orbits in the motion. 
The deformation destroys the infinite number of periodic orbits in each 
family and leaves just an equal number of stable and unstable orbits. The 
stable orbits are surrounded by closed curves (``islands'') which indicate 
the oscillatory motion of nearby 
trajectories around the stable periodic 
orbit. The simplest example in Fig.~4B are the two islands around the 
stable (short) diametral orbit which collides with $\sin(\chi) = 0$ 
at $\phi =  90^{\circ}$. The unstable orbits generate regions of chaotic motion nearby them, 
which correspond to the grainy structureless regions of the SOS. The most 
visible example in Fig.~4B extends around the period-2 islands, reaching 
the $\sin(\chi) = 0$ axis at the location of the unstable (long) diametral orbit 
(which has $\sin(\chi) = 0$ and $\phi = 0,\, 180^{\circ}$). The ``bow-tie'' modes which we have 
focused on in the previous sections would correspond to a four-bounce 
orbit centered on the diametral orbit around $\phi=  90^{\circ}$, but no such orbit 
exists at this low deformation.

To confirm that the relevant resonances at this low deformation are of the 
``whispering gallery'' type, one can generate a phase-space representation 
of the intensity pattern of Fig.~3B, called the 
Husimi function(35). 
For the resonance with deformation $\epsilon = 0.06$ shown in Fig.~3B, this 
function (shown in Fig.~4E) demonstrates that the ray motion corresponding 
to this state is spread out in the large chaotic region just mentioned. 
Note that since the chaotic region extends through $\sin(\chi) = 0$ an orbit in 
this region of phase space will eventually change its sense of rotation 
and is not a ``whispering gallery'' orbit in the familiar sense. However the 
Husimi function of Fig.~4E does not have support near $\sin(\chi) = 0$ 
indicating that escape occurs before this reversal of circulation can 
happen; hence the corresponding real-space intensity pattern (Fig.~3B) 
does have a minimum in the center bounded by an approximate caustic. Note 
further that this orbit lies entirely outside the influence of the central 
diametral orbit and collides with all regions at the  
boundary, thus it may 
reasonably be termed a ``chaotic whispering gallery'' orbit.

At a deformation $\epsilon = 0.10$, the ``bow-tie'' orbit appears at a 
period-doubling bifurcation (33) of the stable diametral orbit. In this 
case, it is a non-generic period-doubling bifurcation (36) in which a new 
stable orbit of twice the period is born (the ``bow-tie''), while 
simultaneously two new unstable, V-shaped, period-2 orbits (``birds'') are 
born (see Fig.~5). Such period-doubling bifurcations are well-understood 
and can be described quantitatively within the general formalism of 
non-linear hamiltonian dynamics (33). However, in this case one can also 
use a more elementary argument familiar from resonator theory. The stable 
(vertical) diametral orbit supports standard Gaussian Fabry-Perot modes 
which are too low-Q to lase in this structure, due to the relatively low 
reflectivity at normal incidence. When the radii of curvature at the two 
contact points of this orbit 
\onecolumn
\begin{figure}[bt]
\hbox{\psfig{figure=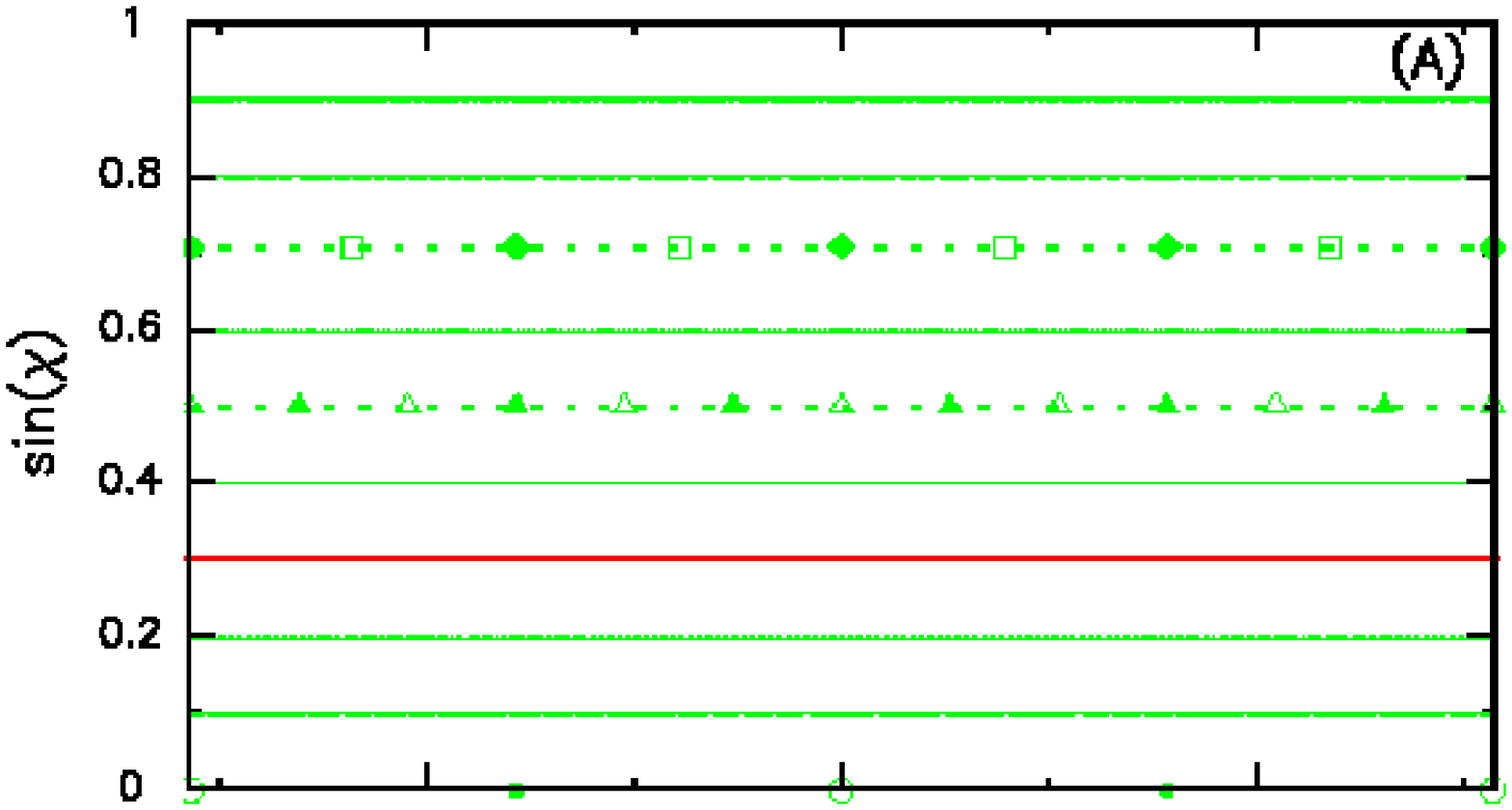,width=8.8cm}
\hspace*{0.25 cm}\psfig{figure=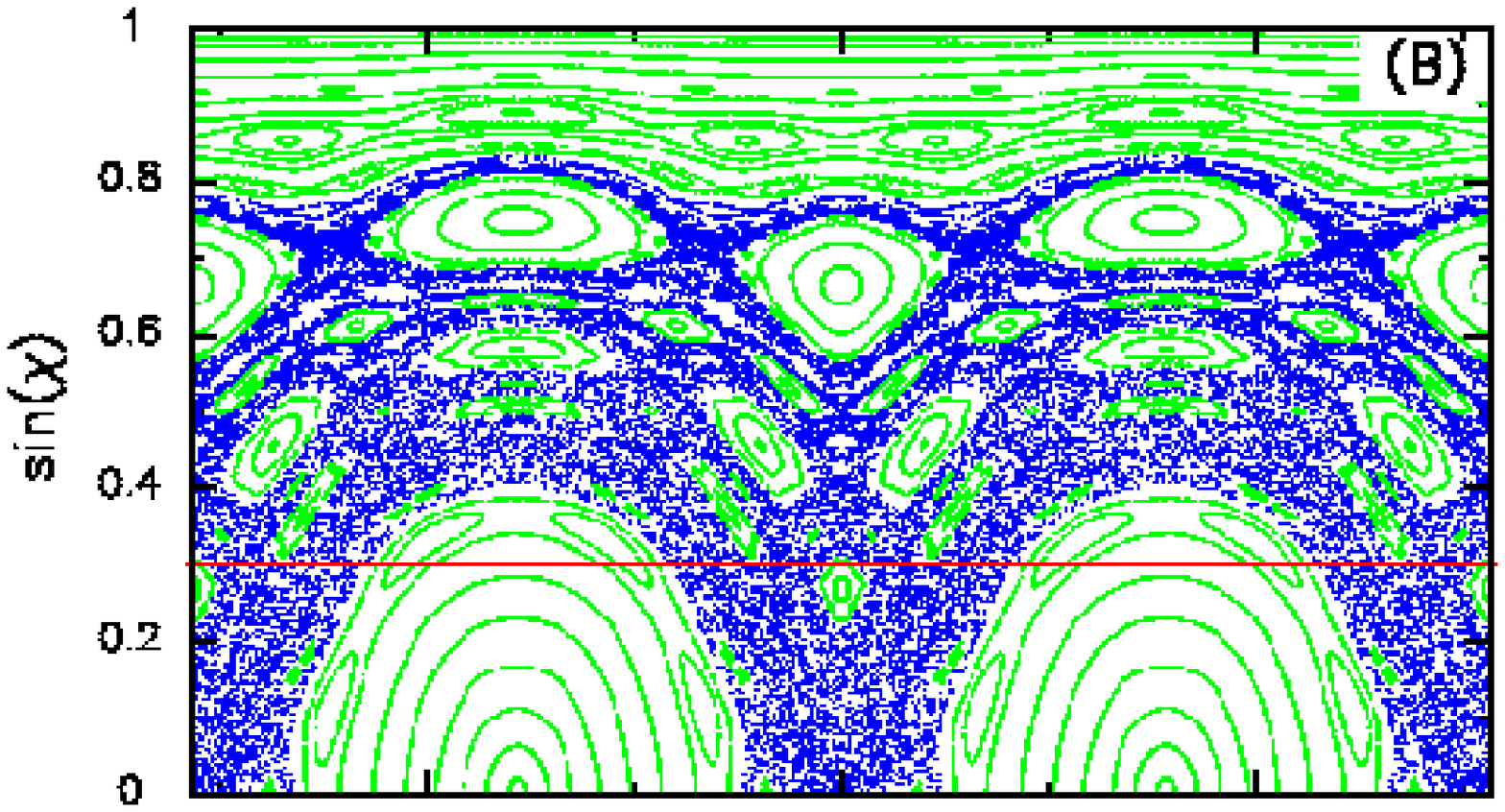,width=8.8cm}}\vspace*{2mm}
\hbox{\psfig{figure=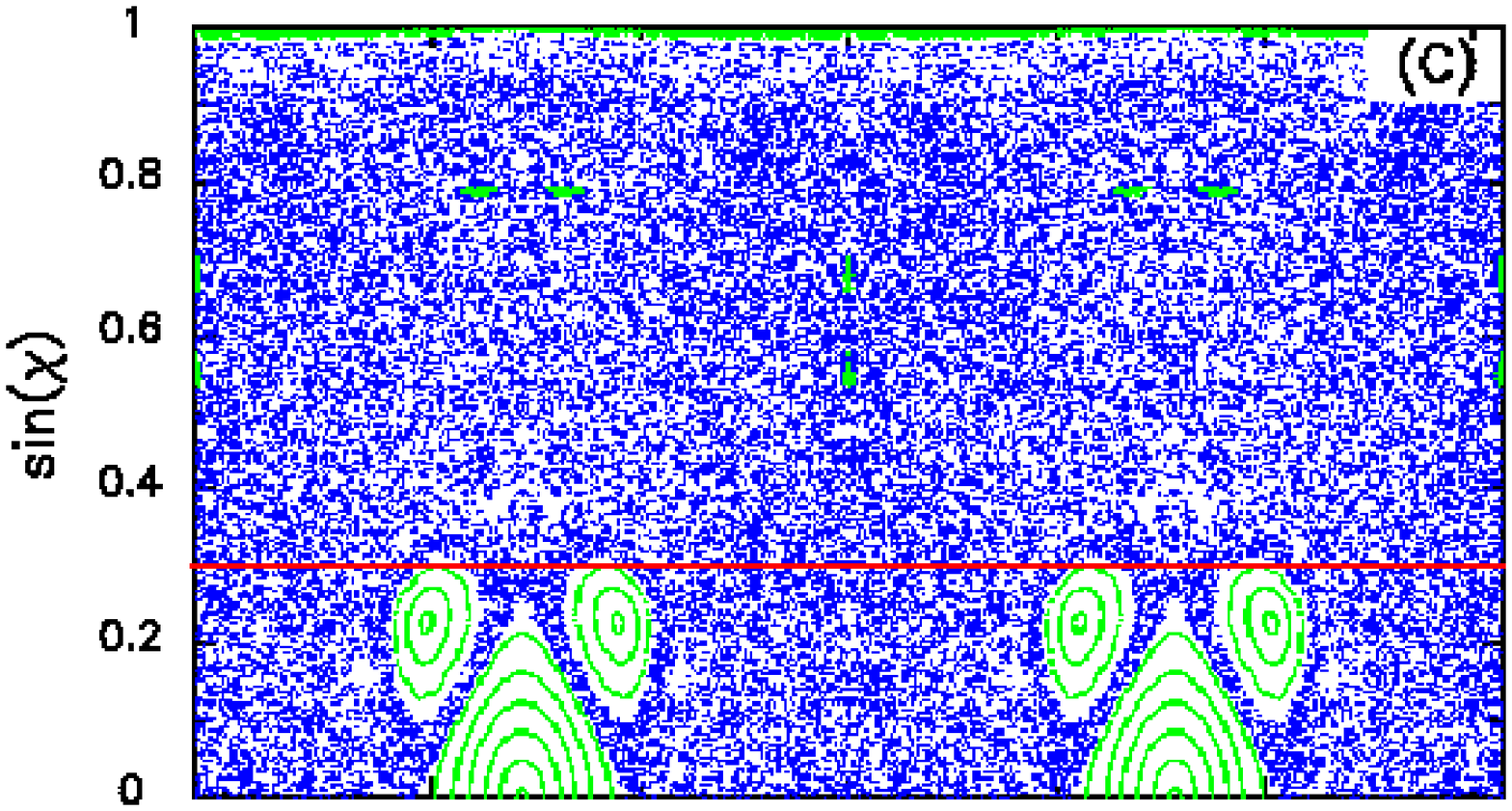,width=8.7cm}
\hspace*{0.4 cm}\psfig{figure=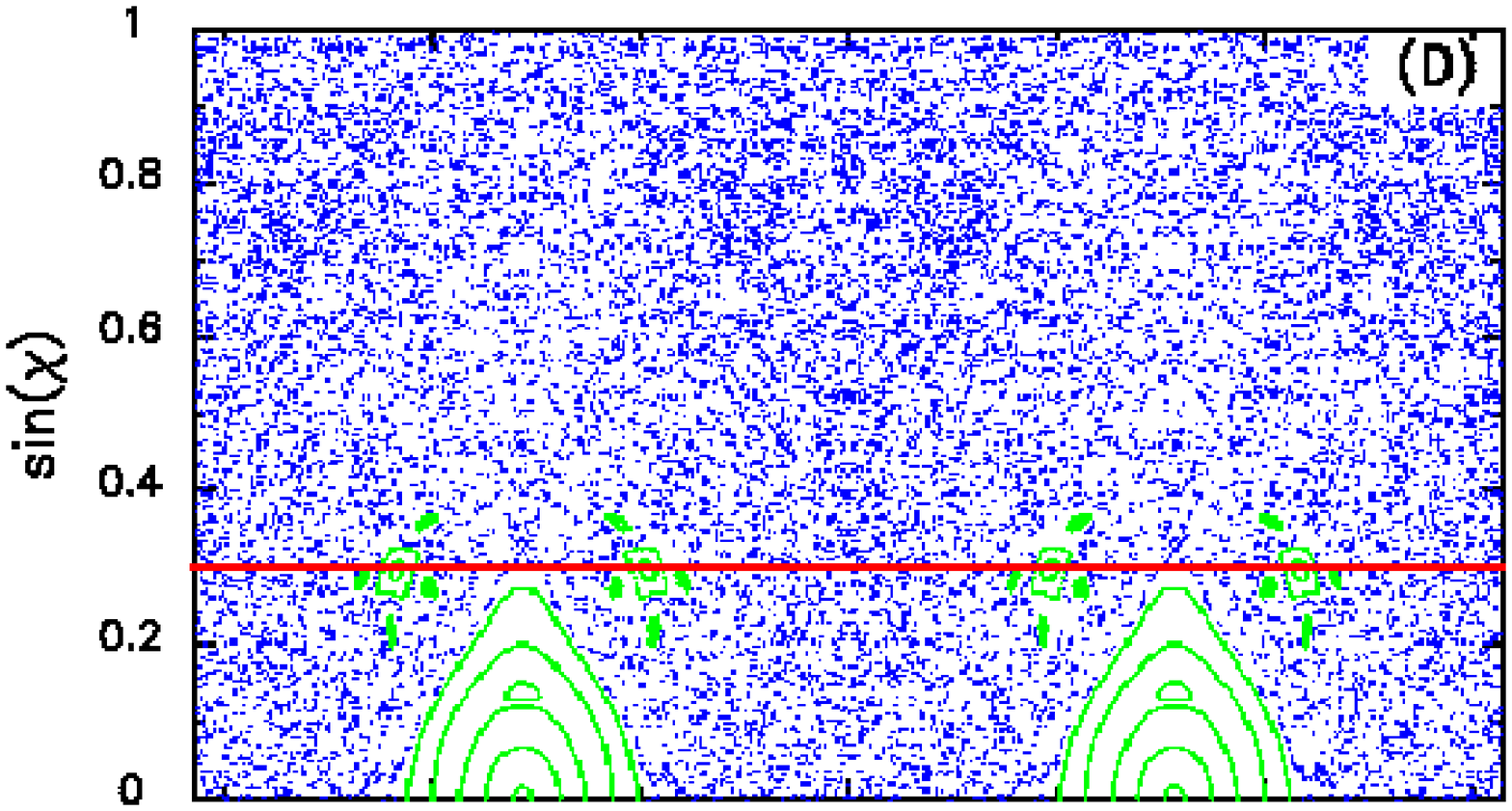,width=8.7cm}}\vspace*{2mm}
\hbox{\psfig{figure=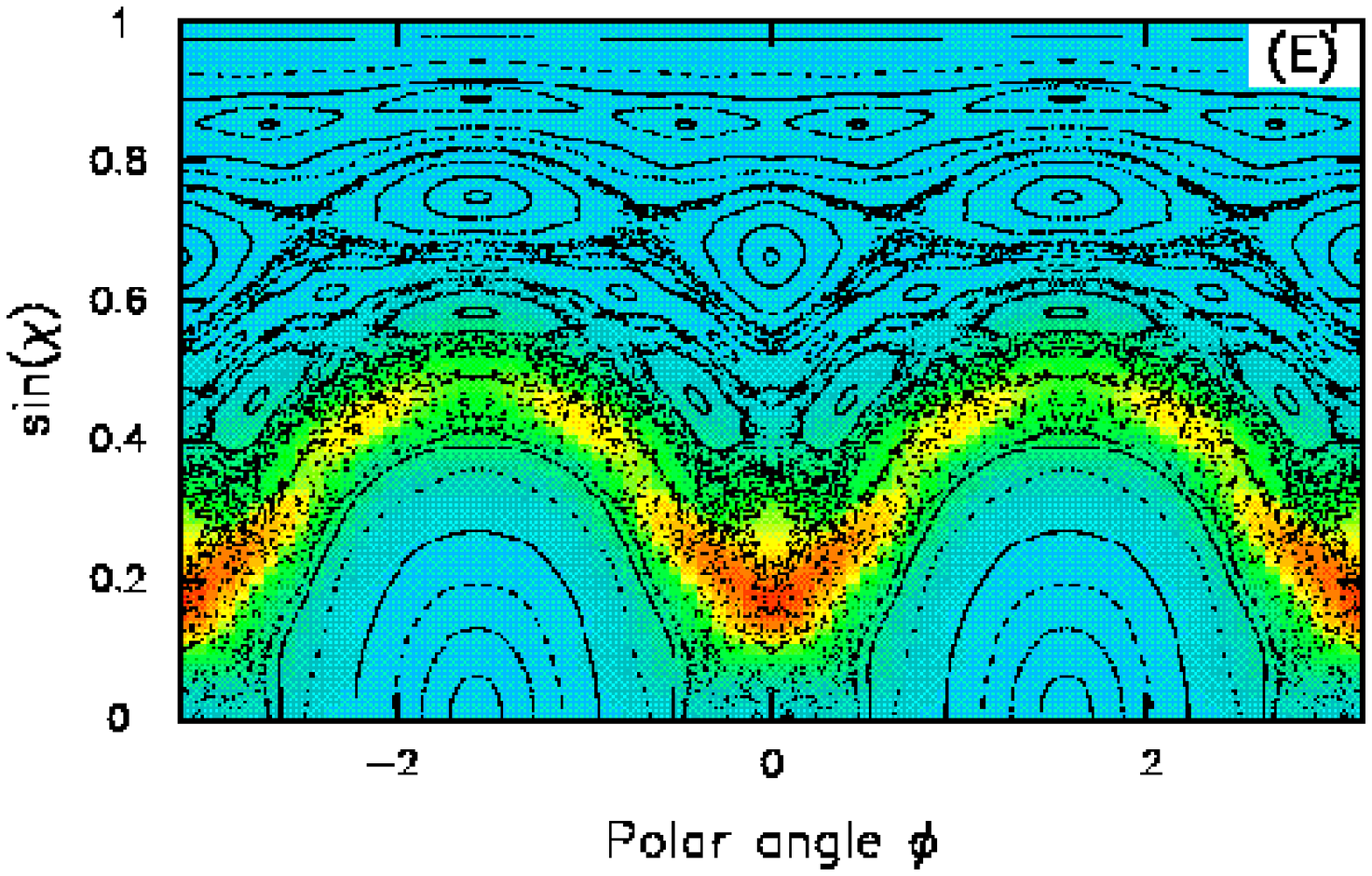,width=8.7cm}
\hspace*{0.4 cm}\psfig{figure=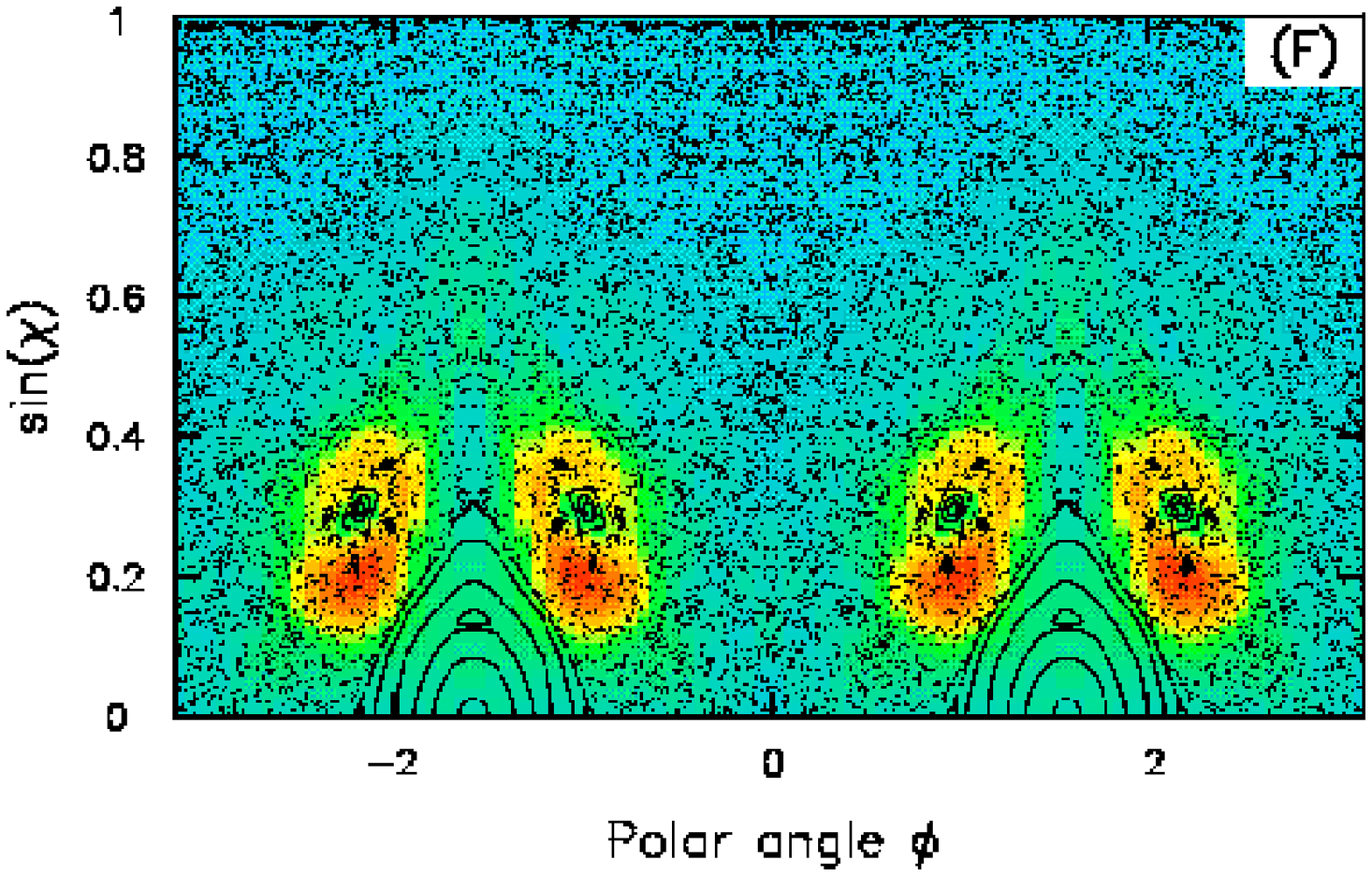,width=8.7cm}}

\footnotesize\noindent 
Figure 4:(A to D) 
Poincar{\'e} surface of section representing the motion of an 
ensemble of rays in phase space for the flattened quadrupolar billiard, 
neglecting the possibility of refractive escape. Regions of stable 
or regular motion are indicated in green, and regions of chaotic motion 
are indicated in blue. $\phi$ is the polar angle in radian as defined in
Fig.~3C, right inset. ({\bf A}) The undeformed (circular) cylinder. Each 
trajectory collides with the boundary at a fixed value of the angle of 
incidence, $\sin(\chi)$, sometimes closing on itself and forming a periodic 
orbit, otherwise passing arbitrarily close to any point on the boundary 
and forming a line in the surface of section. Several members of the 
infinite families of period-2 (circles), period-3 (triangles), and 
period-4 (squares, diamonds) orbits are shown. These members survive to 
nonzero deformation, filled symbols represent orbits that will give 
stable islands, empty symbols those that will be unstable and generate 
regions of chaos. The red line represents the escape condition, 
$\sin(\chi) = 1/n$; in the true resonator rays below that line would rapidly 
escape by Fresnel scattering. ({\bf B}) The phase space for deformation 
$\epsilon = 0.06$, corresponding to the calculation of Fig.~3B. The two major 
islands at polar angle $\phi =\pm\pi/2\,(\pm 90^{\circ})$ 
correspond to motion around the 
stable diametral orbit. Just above these islands is the chaotic region 
generated by the unstable diametral orbit at 
$\phi =\pm\pi\,(\pm 180^{\circ}$).   This region contains the chaotic 
``whispering gallery'' modes. ({\bf C}) For $\epsilon = 0.125$, 
somewhat after the bifurcation of the diametral orbit 
(at $\epsilon = 0.10$) that 
gives rise to the ``bow-tie'' orbits which are seen clearly as the four 
islands at $\sin(\chi) = 0.22$. These islands are sufficiently below the 
critical line that the corresponding modes would be too short-lived to 
lase. Note that there are four symmetric islands for negative $\sin(\chi)$ which 
are not shown; any one ``bow-tie'' orbit only visits four of the islands, 
two with positive and two with negative $\sin(\chi)$, but the same path is 
traced in either case. ({\bf D}) The behavior at $\epsilon = 0.15$ corresponding to the 
data and calculation of Fig. 3, C and D. Now the ``bow-tie'' islands have 
moved up to the critical line, increasing the lifetime of the 
corresponding modes and allowing them to lase. ({\bf E}) Husimi function 
corresponding to the resonance of Fig.~3B at $\epsilon = 0.06$; this function 
clearly represents a chaotic ``whispering gallery'' state localized in the 
chaotic region. The Husimi function translates the real-space electric 
field intensity pattern into a probability density in phase-space. The 
resulting function is illustrated by a color scale, where red is high 
intensity. 
({\bf F}) The relation between the highly directional resonator mode 
shown in Fig.~3D and the islands shown in Fig.~4D can be demonstrated by 
means of the Husimi function shown in Fig.~4F. This function is centered 
on the ``bow-tie'' islands. The minimum at the very center indicates that 
this ``bow-tie'' mode has an oscillatory motion transverse to the ``bow-tie'' 
path; this is consistent with the intensity pattern of Fig.~3D which 
exhibits four transverse oscillations.
\end{figure}
\twocolumn
\begin{figure}
\hspace*{1.5cm}\psfig{figure=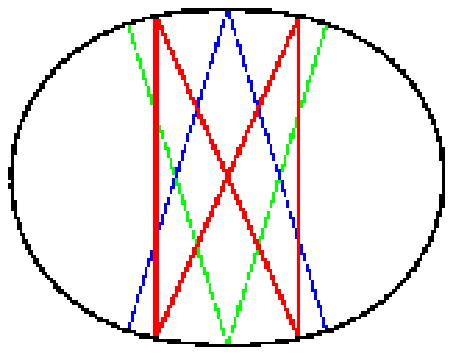}
	
\footnotesize\noindent 
Figure 5:
The three orbits born at the period-doubling bifurcation of the 
stable diametral orbit, the stable ``bow-tie'' (red) and the two unstable 
V-shaped ``birds'' (blue, green). This occurs at a deformation 
$\epsilon = 0.10$. The 
birds, being unstable, will not generate long-lived resonator modes; 
however, the stable ``bow-tie'' generates modes with directional properties 
and spectral spacing in excellent agreement with the experiment as 
discussed above. A key feature of the ``bow-tie'' is that it does not exist 
until the resonator is substantially deformed so that the confocal 
condition is reached for the stable diametral orbit, as discussed in the 
text.
\end{figure}
\noindent
become equal to the distance between them 
(the minor axis $L$) we reach the confocal condition (37) at which 
marginally stable families of ``bow-tie'' and V-shaped orbits, all of length 
$4L$, come into existence. For the flattened quadrupole this occurs at 
$\epsilon = 0.10$. For slightly larger deformations these orbits leave the vicinity 
of the diametral orbit and do not correspond to small deformations of 
diametral orbits. Such orbits are not typically discussed in Fabry-Perot 
theory (38). But here, because the boundary creates a full $180$ degree 
``mirror'' with a reflectivity which increases at oblique incidence, the 
modes associated with the remaining stable ``bow-tie'' orbit are higher-Q 
than the simple Fabry-Perot modes and can lase when the latter do not. 
Because they require a doubling of the radius of curvature at the minor 
axis, they do not exist at small deformations. 

The SOS for $\epsilon = 0.125$ shown in Fig.~4C is taken just after the bifurcation 
of the diametral orbit, showing the emergence of the stable ``bow-tie'', 
which has the feature that its angle of incidence is the same for all four 
bounces (see caption to Fig.~4C). Note, however, that for this deformation 
$\sin(\chi) = 0.22$, which is still well below the critical angle.

When the deformation is further increased to $\epsilon \ge 0.14$ the ``bow-tie'' orbit 
has moved upward in the SOS so that it is centered near the critical angle 
(see Fig.~4D). The reflectivity of the corresponding modes will increase 
to a value comparable to the ``whispering-gallery'' modes at the 
same $\sin(\chi)$ 
in the circle. Therefore we expect a turn-on of the laser emission from 
this mode. Note that the ``bow-tie'' orbit now represents the only large 
stable island at or above the critical angle in the SOS. Moreover, it is 
now well separated in phase space by a chaotic region from the fundamental 
diametral orbit from which it originated. The plot in Fig.~3D shows the 
high-intensity regions concentrated on this orbit. In Fig.~4F, we show the 
phase-space projection of this mode, which is concentrated in the vicinity 
of the islands corresponding to the ``bow-tie'' orbit.

The modes corresponding to the ``bow-tie'' orbit are not simply higher order 
transverse Fabry-Perot modes; the latter would correspond to quantized 
oscillations within the island around the diametral orbit. Moreover, as 
noted above, the ``bow-tie'' orbit is rather different from the 
``whispering-gallery'' orbits because the sense of rotation of the ``bow-tie'' 
orbit is not constant; it represents a librational rather than a 
rotational motion.

The existence and stability of the ``bow-tie'' orbit is relatively 
insensitive to the precise shape of the boundary, so we expect these modes 
to be generic to deformed cylindrical resonators. For the flattened 
quadrupole, the stable ``bow-tie'' exists in the range of deformations from 
$\epsilon = 0.10$ to $\epsilon\approx 0.23$. Its directions of peak emission, though, are 
sensitive to the precise shape of the resonator; the degree of sensitivity 
will be the subject of future studies (38). Nevertheless, reasonable 
agreement between theory and experiment has been obtained for the 
far-field directionality using the flattened quadrupolar shape (Fig.~3C).

As noted above, for the range of deformations at which the stable 
``bow-tie'' orbit exists, it represents the only substantial islands of 
stability in the region of phase-space close to the critical value of 
total internal reflection; thus it is difficult to find a competitive 
mechanism for the highly directional modes we observe. At lower 
deformations, other librational modes exist and may be important in the 
cross-over from ``whispering gallery'' to ``bow-tie'' emission.

Highly directional emission from low refractive index resonators was 
discussed in earlier theoretical work by several of the authors, and 
tested in experiments on lasing dye-jets (12). However, the origin of 
directionality at high deformations in the high refractive index 
resonators discussed in the present paper is qualitatively different from 
the mechanism studied in this earlier work. In resonators with indices of 
refraction $n < 2$, the escape line corresponding to $\sin(\chi) =
1/n$ is much 
higher in the surface of section. Therefore a ray escaping from a 
``whispering-gallery'' mode must traverse a much smaller fraction of the 
chaotic sea to escape. It has been shown (11,12) that in this case the 
motion is not pseudo-random, and highly directional emission from near the 
points of highest curvature results. However, in the high-index materials 
of the present work, it is necessary to reach much lower angles of 
incidence within the resonator to escape and we now find that the escape 
direction for rays starting far from the critical angle is effectively 
random, at least for the deformations where the ``bow-tie'' orbit
exists.
This is demonstrated by the chaotic scattering map shown in Fig.~6. As is 
explained in the caption to Fig.~6, this map suggests strongly that highly 
directional modes of the ``whispering-gallery'' type are not easily achieved 
at high deformations in such resonators
\begin{figure}[bt]
\psfig{figure=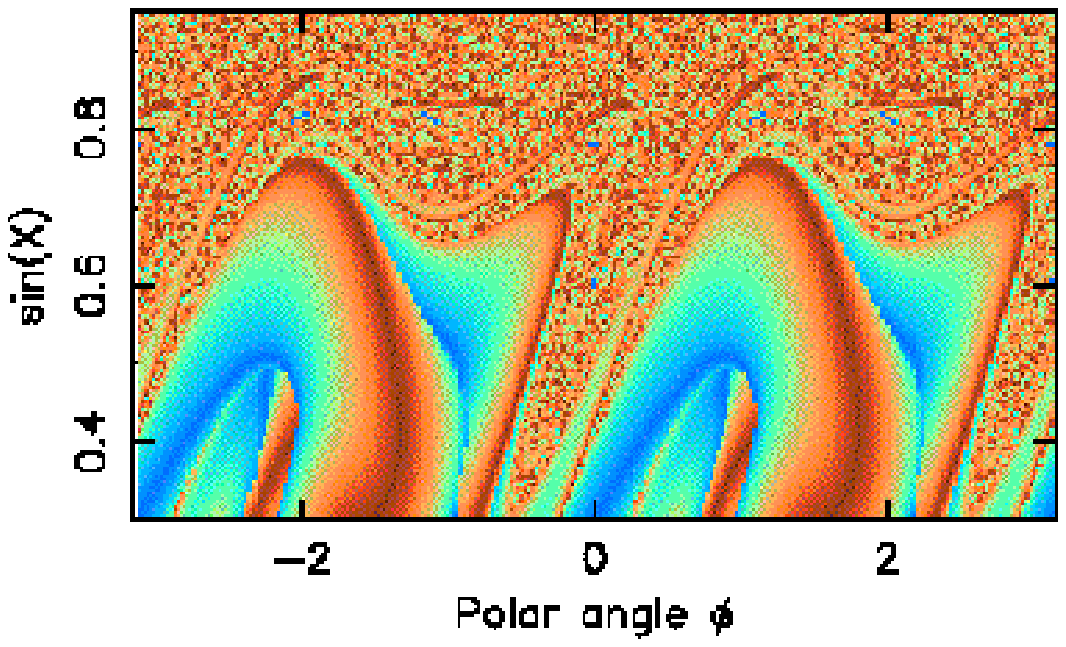,width=7.5cm}

\footnotesize\noindent 
Figure 6:
Color representation of the directionality of escaping rays for 
the phase space of the flattened quadrupole at $\epsilon = 0.15$. 
Blue represents 
initial conditions leading to escape into the far-field at $0^{\circ}$
and  red at 
$90^{\circ}$. Note the clear demarcation between a pseudo-random region with 
rapidly fluctuating escape direction for initial angles $\sin(\chi) > 0.70$ and 
a regular region where the escape direction varies smoothly and relatively 
slowly. In the regular region, which primarily corresponds to librational 
motion, escape is so rapid that chaos cannot fully develop; in contrast, 
initial conditions in the ``whispering gallery'' region above $\sin(\chi) = 0.70$ 
must traverse more of the chaotic sea and cannot generate highly 
directional escape. This behavior is completely different from low-index 
resonators with the same magnitude of deformation (11). The regular region 
can generate directional emission, but only for states localized by 
islands, such as the ``bow-tie'' states.
\end{figure}
\noindent  made from semiconductor materials, 
although such modes exist and dominate the lasing properties at the same 
range of deformations for lower index materials. Conversely, modes such as 
the ``bow-tie'' resonance, which are related to librational orbits, all 
reside well below $\sin(\chi) = 0.5$ and as such would experience too little 
reflectivity from the boundary to reach laser threshold in low-index 
materials. Note finally that the ``bow-tie'' modes are confined away from 
the points of highest curvature in the resonator and thus display a 
minimum in the near-field intensity at these points (Fig.~3D) in contrast 
to the ``whispering gallery'' modes (Fig.~3B) which have high intensity in 
the near-field at these points. Therefore the two types of modes should be 
easily distinguishable if near-field measurements could be made. This is a 
demanding task for lasers in the mid-infrared region of the spectrum and 
will be the subject of future research. Finally, it should be emphasized 
that there is no fundamental reason that such resonators should not be 
equally effective as microcavities at visible wavelengths. 

\vspace*{0.5 cm}\begin{center}{\large\bf REFERENCES AND NOTES:}\\
\rule{6.8cm}{1pt}\end{center}
\begin{enumerate}
\item{Y.~Yamamoto and R.~E.~Slusher, Physics Today {\bf 46}, 66 (1993).}
\item{S.L.McCall, A.F.J.Levi, R.E.Slusher, S.J.Pearton, and 
R.~A.~Logan, Appl.~Phys.~Lett.~{\bf 60}, 289 (1991); H.~Deng, Q.~Deng, and 
D.~G.~Deppe, Appl.~Phys.~Lett.~{\bf 69}, 3120 (1996); T.~Baba, IEEE 
Select.~Top.~Quantum Electron.~{\bf 3}, 808 (1997).}
\item{	A.~F.~J.~Levi, R.~E.~Slusher, 
S.~L.~McCall, S.~J.~Pearton, and W.~S.~Hobson, 
Appl.~Phys.~Lett.~{\bf 62}, 2021 (1993).}
\item{	S.-X.~Qian, J.~B.~Snow, H.-M.~Tzeng, and R.K.~Chang, Science {\bf 231}, 486 (1986).}
\item{	H.~Yokoyama and K.~Ujihara, Eds., {\em Spontaneous Emission and Laser 
Oscillation in Microcavities} (CRC Press, Boca Raton, 1995).}
\item{	M.F.~Crommie, C.P.~Lutz, D.M.~Eigler, Science {\bf 262}, 218
(1993). 
C.~M.~Marcus, A.~J.~Rimberg, R.~M.~Westervelt, P.~F.~Hopkins, and 
A.~C.~Gossard, Phys.~Rev.~Lett.~{\bf 69}, 506 (1992).}
\item{	A.~Kudrolli, V.~Kidambi, and S.~Sridhar, Phys.~Rev.~Lett.~75, 822 
(1995); H.Alt et al., Phys.~Rev.~Lett.~{\bf 74}, 62 (1995); S.Sridhar, 
Phys.~Rev.~Lett.~{\bf 67}, 785 (1991); 
J.~Stein and H.~J.~St{\"o}ckmann, Phys.~Rev.~Lett.~{\bf 68}, 
2867 (1992); H.~D.~Gr{\"a}f et al.~Phys.~Rev.~Lett.~{\bf 69}, 1296 (1992).}
\item{	A.~F.~J.~Levi et al., Appl.~Phys.~Lett.~{\bf 62}, 561 (1993); D.~Y.~Chu et 
al., Appl.~Phys.~Lett.~{\bf 65}, 3167 (1994).}
\item{	B.-J.~Li and P.-L.~Liu, IEEE J.~Quantum.~Electron.~{\bf 33}, 791 (1997).}
\item{J.~Faist et al., Science {\bf 264}, 553 (1994); F.~Capasso, J.~Faist, 
C.~Sirtori, A.~Y.~Cho, Solid State Commun.~{\bf 102}, 231 (1997).}
\item{J.~U.~N{\"o}ckel and A.~D.~Stone, Nature {\bf 385}, 45 (1997).}
\item{	J.~U.~N{\"o}ckel, A.~D.~Stone, G.~Chen, H.~Grossman, and R.~K.~Chang, Opt.~Lett.~{\bf 21}, 1609 (1996); A.~Poon and R.~K.~Chang, unpublished material.}
\item{J.~U.~N{\"o}ckel and A.~D.~Stone, in {\em Optical Processes in Microcavities}, 
R.~K.~Chang and A.~J.~Campillo, Eds.~(World Scientific Publishers, 
Singapore, 1995), chap.11.}
\item{	J.~U.~N{\"o}ckel, A.~D.~Stone, and R.~K.~Chang, Opt.~Lett.~{\bf 19}, 1693 (1994).}
\item{	W.~T.~Silfvast, {\em Laser Fundamentals}  (Cambridge University Press, 
Cambridge, 1996).}
\item{	H.~Liu, S.-H.~Zhou, and Y.~C.~Chen, Opt.~Lett.~{\bf 23}, 451 (1998).}
\item{	J.~Faist et al., Appl.~Phys.~Lett.~{\bf 69}, 2456 (1996); C.~Gmachl et al., 
IEEE J.~Quantum.~Electron.~{\bf 33}, 1567 (1997).}
\item{	The electric field vector is oriented vertical to the laser plane, 
therefore light-coupling is prohibited in this direction.}
\item{	J.~Faist et al., Appl.~Phys.~Lett.~{\bf 68}, 3680 (1996); J.~Faist et al., 
IEEE J.~Quantum Electron.~{\bf 34}, 336 (1998).}
\item{	The waveguide core contains 25 cascaded stages each consisting of a 
so-called ``three-well vertical transition'' active region and an electron 
injector as described in (19). The waveguide core is sandwiched between 
two waveguide cladding layers. Each cladding is composed of three 
sub-layers: a low doped GaInAs layer (Si doping level $n = 
2\times10^{17}$ cm$^{-3}$, 
thickness $d = 350$ nm) adjacent to the active material, an inner low doped 
AlInAs layer ($n = 2\times 10^{17}$ cm$^{-3}$, $d = 300$ nm, and $n = 
3\times 10^{17}$ cm$^{-3}$, $d = 400$ 
nm) and an outer highly doped AlInAs layer ($n = 7\times 10^{18}$ 
cm$^{-3}$, $d = 1000$ nm), 
which serves as a plasmon confinement layer (21).At the upper cladding 
hetero-interface between the GaInAs and AlInAs layers a 2D electron gas 
(2DEG) (22) is formed by highly doping a thin slice ($n = 5\times 
10^{18}$ cm$^{-3}$, $d = 
8$ nm) of the AlInAs layer close to the interface. This together with a 
highly doped final cap-layer (Sn: $n = 1\times 10^{20}$ cm$^{-3}$, $d = 
100$ nm) of the 
structure facilitates lateral current spreading.}
\item{	C.~Sirtori et al., Appl.~Phys.~Lett.~{\bf 66}, 3242 (1995).}
\item{	L.~Pfeiffer, E.~F.~Schubert, K.~W.~West, Appl.~Phys.~Lett.~{\bf 58}, 2258 
(1991).}
\item{	The samples are strongly agitated and etched in an aged solution of 
HBr : HNO$_3$ : H$_2$O $= 1 : 1 : 10$ for several minutes at room temperature.}
\item{	The rim of the top contact pad has the same distance from the edge of 
the laser disk (with a small variation of $\approx  5 \%$) for each laser, in all 
directions $\phi$, and at all deformations $\epsilon$. The cylinder lasers have been 
fabricated with their long diameter oriented in $0^{\circ}, \,45^{\circ}$, 
and $90^{\circ}$ relative 
to the major orientation of the semiconductor crystal. Finally the 
processing leaves the surface clean without evident sources of surface 
roughness scattering. These precautions ensure that there is no additional 
directionality introduced into the system other than through the flattened 
quadrupolar shape.}
\item{	The maximum peak power is a widely accepted valid measure of the 
power performance of semiconductor lasers since it refers to the actual 
useful power. In the QC-laser the peak optical power is reached when the 
material gain decreases mainly due to two effects: the loss of optimum 
alignment of the ground state of the injector with the upper level of the 
laser transition with increasing voltage (39) and thermal population of 
the lower laser level with increasing current density. Therefore the peak 
power is primarily insensitive to size-effects (when normalized to the 
unit area or volume).}
\item{	There is no strong theoretical basis for an ``exponential'' power 
increase, in particular since it is based on several different effects and 
is far-field dependent. Nevertheless, we chose the term ``exponential'' 
increase, since it represents the data qualitatively well.}
\item{	The continuity conditions at the boundary cannot be satisfied for real 
values of the wavevector if there is no incident wave, so one looks for 
solutions with complex wavevectors. It can be shown that the real part of 
these solutions gives the wavevector at which scattering resonances would 
occur for a wave incident from infinity, while the imaginary part gives 
the width ($Q$-value) of the resonance. In a scattering experiment the 
measured intensity in the far-field has contributions both from the 
resonant scattering and the incident beam, whereas in lasing emission only 
the resonant emission is present. Hence it is the intrinsic emission 
pattern of the quasi-bound state which is measured in the experiments 
reported above and it is this quantity which we plot in Fig.~3, B and 
D. See (13), chap.~1 for a detailed discussion.}
\item{	M.V.Berry, {\em The Bakerian Lecture}, Proc. Roy. Soc.~A {\bf 413}, 183 (1987).}
\item{	M.C.Gutzwiller, {\em Chaos in Classical and Quantum Mechanics }
(Springer-Verlag, New York, 1990).}
\item{	L.~Reichl, {\em The Transition to Chaos in Conservative Classical Systems: Quantum Manifestations} (Springer-Verlag, New York, 1992).}
\item{	W.~H.~Miller, J.~Chem.~Phys.~{\bf 56}, 38 (1972); P.~Gaspard and S.~A.~Rice, 
J.~Chem.~Phys.~{\bf 90}, 2225 (1989).}
\item{	M.~V.~Berry, Eur.~J.~Phys.~{\bf 2}, 91 (1981).}
\item{	B.~R.~Johnson, J.~Opt.~Soc.~Am.~{\bf 10} , 343 (1993).}
\item{	V.~F.~Lazutkin, {\em KAM Theory and Semiclassical Approximations to 
Eigenfunctions }(Springer Verlag, Berlin, 1993).}
\item{	The ``Husimi-function'' is the squared overlap of the interior electric 
field with a minimum-uncertainty wavepacket centered on a given point in 
the surface of section. It may be roughly interpreted as a phase space 
probability density for the photons in the mode. A precise definition is 
given in P.~LeBoeuf and M.~Saraceno, J.~Phys.~A: Math.~Gen.~{\bf 23}, 1745 
(1990).}
\item{	In a generic period-doubling bifurcation the shorter orbit goes 
unstable as a new stable orbit with twice the period is born. Here, due to 
the symmetry, the shorter (diametral) orbit just reaches marginal 
stability, the three orbits described in the text are born, and the 
diametral orbit immediately restabilizes. This is consistent with the 
Poincar{\'e} index theorem since an even number of stable and unstable fixed 
points are created in this process.}
\item{	B.~E.~A.~Saleh, M.~C.~Teich, {\em Fundamentals of Photonics} (Wiley, New 
York, 1991)}
\item{	We may regard the ``bow-tie'' resonances as associated with a 
four-mirror resonator defined by the tangents to the points of contact of 
the ``bow-tie'' orbit. Some general properties of these modes can be derived 
from this point of view, which will be presented elsewhere.}
\item{	C.~Sirtori et al., IEEE J.~Quantum Electron., in press}
\item{We are grateful to A.~Tredicucci for useful 
discussions. E.~E.~N., J.~U.~N., and A.~D.~S. gratefully 
acknowledge support from the Aspen Center for Physics for part of this work. 
The work performed at Bell Laboratories was supported in 
part by DARPA (Defence Advanced Research Project Agency)-U.~S. 
Army Research Office under Contract No. DAAH04-96-C-0026.  
The work performed at Yale has been supported in part by NSF grant 
PHY9612200.}\\[2ex]
23 February 1998; accepted 21 April 1998
\end{enumerate}

\end{document}